# Hilbert Transform based Quadrature Hybrid RF Photonic Coupler via a Micro-Resonator Optical Frequency Comb Source

Thach G. Nguyen,[1*] Mehrdad Shoeiby,[1] Sai T. Chu,[2] Brent E. Little,[3] Roberto Morandotti,[4] Arnan Mitchell,[1,5] and David J. Moss[1]

[1]*Schoof of Electrical and Computer Engineering, RMIT University, Melbourne, VIC 3001, Australia*
[2]*Department of Physics and Materials Science, City University of Hong Kong*
[3]*Xi'an Institute of Optics and Precision Mechanics, CAS, Xi'an, China PRC*
[4]*INSR – Énergie,Matériaux et Télécommunications, 1650 Blvd Lionel Boulet, Varennes (Québec), J3X1S2, Canada*
[5]*ARC Centre of Excellence for Ultrahigh bandwidth Devices for Optical systems (CUDOS)*
[*]*thach.nguyen@rmit.edu.au*

**Abstract:** We demonstrate a photonic RF Hilbert transformer for broadband microwave in-phase and quadrature-phase generation based on an integrated frequency optical comb, generated using a nonlinear microring resonator based on a CMOS compatible, high-index contrast, doped-silica glass platform. The high quality and large frequency spacing of the comb enables filters with up to 20 taps, allowing us to demonstrate a quadrature filter with more than a 5-octave (3 dB) bandwidth and an almost uniform phase response.

## 1. Introduction

Many applications, including radar mapping, measurement, imaging as well as the realization of advanced modulation formats for digital communications, require the generation, analysis and processing of analogue RF signals. In these applications, both the amplitude and phase of the signals are critically important. In order to access such information, it is often necessary to perform a uniform quadrature-phase shift ($\pm 90^{o}$) of the constituent RF frequencies over the entire bandwidth of interest. The process of obtaining in-phase and quadrature-phase components of a signal can be achieved via a Hilbert transform, and for RF signals this is often achieved using a so-called 'hybrid coupler' [1]. While microwave hybrid coupler technology is very mature, the performance of these components is often degraded by large amplitude and phase ripple [2-4]. In addition, it is difficult to achieve wide band operation using electronic circuits. In many applications, especially those involving radar and early warning receivers in electronic warfare systems, the capacity to process signals over a multi-octave spectrum from below 1 GHz to 20 or even 40 GHz is generally required [5]. Electronic approaches to microwave signal processing often need banks of parallel systems to cover such a wide spectral bandwidth.

Photonic implementations of Hilbert transformers have been demonstrated that have achieved very high performance over a very wide bandwidth. Furthermore, these devices can offer immunity to electromagnetic interference, while being compatible with fiber remote distribution systems and parallel processing. There are several approaches to implementing photonic Hilbert transformers, including phase-shifted Bragg grating devices [6-8] or Fourier domain optical processors [9]. However, the performance in terms of bandwidth and low frequency cutoff typically associated with Hilbert transformers based on phase-shifted Bragg gratings has been quite limited due to the difficulty of fabricating grating devices with the necessary stringent requirements [6-8]. To address these challenges in realizing Hilbert transformers, programmable Fourier domain optical processors [9] have been proposed as an approach to greatly reduce amplitude and phase ripple over very broad RF bandwidths. However, even this approach suffers from performance degradation at frequencies below 10 GHz due to typical resolution limits of Fourier domain optical processors [9]. Recently, a tunable fractional temporal Hilbert transformer based on the phase shift at the resonant wavelength of a ring resonator has been demonstrated [10]. Although the Hilbert transformers based on grating or ring resonator structures can provide a phase shift at the center frequency within the device bandwidth, accessing to both the in-phase and quadrature phase components of the same signal is not available.

Transversal filtering is a versatile approach that can achieve a diverse range of signal processing functions. In this approach, the signal is divided into a number of copies (taps) with different weights, each copy having a tap that is delayed in time by incremental amounts. All of the delayed taps are then summed and detected. Each of the taps can be considered a discrete sample of the filter impulse response. By carefully controlling the amplitudes and delays of all taps, different filtering functions can be realized.

Microwave Photonic transversal filters have been reported by several groups [11], including a broadband photonic Hilbert transformer for in-phase and quadrature-phase generation [12, 13] that exhibited low amplitude ripple and very low phase error over a multi-octave bandwidth [12]. The application of this transformer to instantaneous frequency measurement (IFM) has been reported [14-16]. Photonic transversal filtering has typically relied on multiple discrete laser diodes for the multi-wavelength source. However, since the filter bandwidth strongly depends on the number of filter taps (i.e. the number of wavelengths) and therefore the number of laser diodes, this typically results in a very high cost, complexity, energy consumption, and footprint, particularly for systems that are parallelized. Therefore, as an alternative to using individual lasers for each tap, a single component that can generate multiple, and high quality, wavelengths would be highly advantageous.

High quality, large spectral range multi-wavelength sources have typically been based on mode-locked fiber lasers [17], electro-optically generated combs [18] and nonlinear micro-resonators [19]. Of these, integrated optical frequency combs based on the Kerr nonlinearity in nonlinear micro-resonators have displayed significant potential for high performance transversal filtering for RF signal processing. These devices greatly reduce both cost and complexity due to their very compact size as well their suitability for monolithic integration. Significant progress has been made in demonstrating high quality, wide spectrum optical frequency combs [20-25]. Moreover, unlike the combs generated from mode-locked lasers or RF modulation, micro-resonators can achieve very large frequency spacings [24], allowing for the control of the amplitude and phase of individual comb lines via commercially available wave shapers [26]. Indeed, comb sources based on silicon nitride microrings have been successfully used for high performance programmable band-pass RF filter [26].

In this paper, we present a photonic Hilbert transformer based on multi-tap transversal filtering. We use an integrated frequency comb source based on a high Q factor nonlinear microring resonator fabricated in a CMOS compatible, high-index, doped-silica glass (Hydex) waveguide platform [25, 27-30]. The comb source has a wide spectral range with a large frequency spacing of 200 GHz, allowing for the realization of filters with up to 20 taps. We demonstrate an RF quadrature coupler with more than a 5-octave 3dB bandwidth and with a near constant relative phase over the pass band. This represents the first demonstration of a wide band photonic RF Hilbert transformer, or indeed any transversal filter, based on an integrated comb source that has both positive and negative taps.

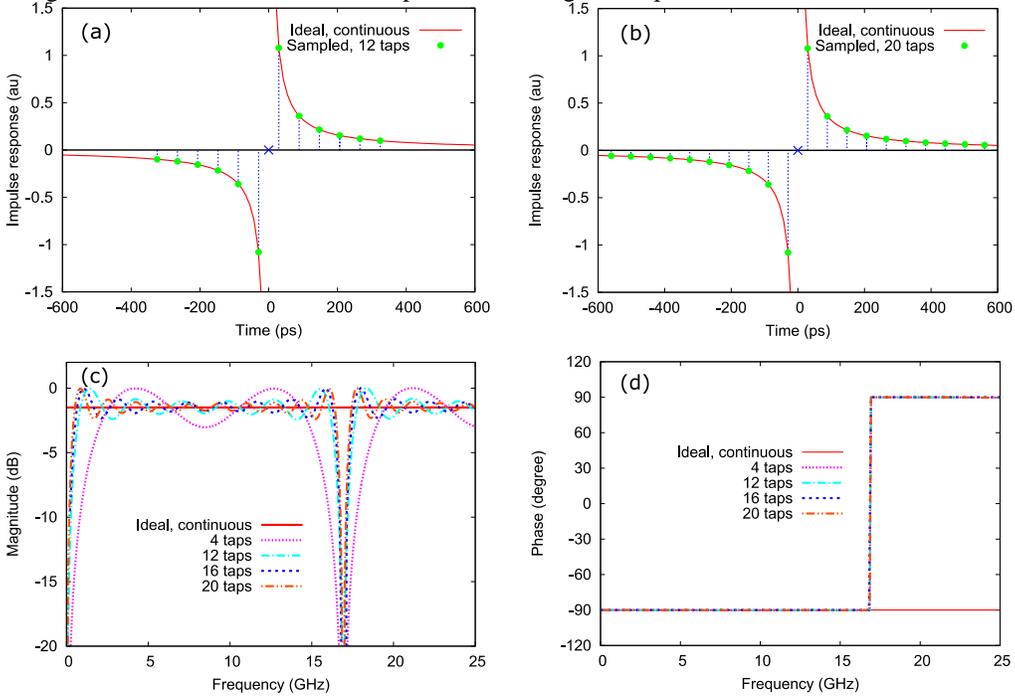

Fig. 1. Working principle of a transversal Hilbert transformer: (a) - (b) Ideal, continuous hyperbolic and discretely sampled impulse response with time spacing Δt = 59 ps. Corresponding (c) amplitude and (d) phase response for the continuous and discrete cases, respectively.

## 2. Photonic Hilbert transformer via transversal filtering

An ideal Hilbert transformer exhibits a constant amplitude frequency response and a ±90° frequency independent phase shift centred around the main (central) frequency. The impulse response of an ideal Hilbert transformer [Fig. 1(a)] is a continuous hyperbolic function $1/(\pi t)$ that extends to infinity in time. In order to realize this impulse response using transversal filtering, the hyperbolic function is truncated and sampled in time by discrete taps [12]. The theoretical RF transfer function of the filter is the Fourier transform of the impulse response. Fig. 1 (c) shows the calculated spectrum associated with the filtered signal amplitude when the impulse response is truncated and evenly sampled. Here, the bandwidth is limited by amplitude ripple, with 'nulls' at frequencies zero and $f_c$. The null frequency $f_c$ is determined by the sample spacing $\Delta t$ as $f_c = 1/\Delta t$. Amplitude ripples and the bandwidth in particular depend strongly on the number of taps, where the bandwidth increases dramatically with the number of sample taps. For example, the 3 dB bandwidth increases from less than 3 octaves with a 4 tap filter to more than 5 octaves when the number of filter taps is increased to 16 or more.

Figure 1(d) shows the calculated frequency response of the filtered signal phase for different numbers of taps in the impulse response. Unlike the amplitude frequency response, when the impulse response function is truncated and sampled, the phase response does not exhibit any ripples from zero up to the null frequency $f_c$. The phase is constant at -90$^0$ regardless the number of samples. At the null frequency, $f_c$, the phase transitions from -90º to 90º.

The simulations presented in Fig. 1 indicate that a band-limited Hilbert transformer can be realized using a transversal filtering method with the tap coefficients set to a hyperbolic function. The bandwidth and pass-band amplitude ripples are determined by the spacing between filter taps and by their number. The Hilbert transform impulse response is an asymmetric function centred at time zero. This has two implications for realizing a practical

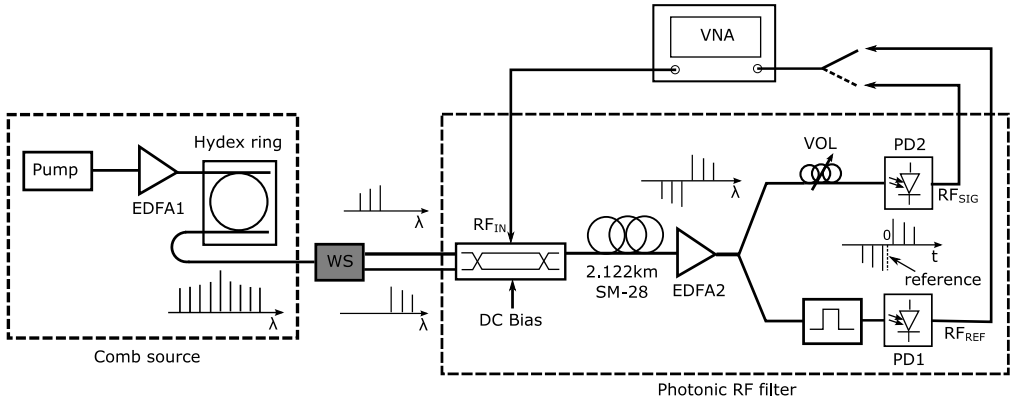

Fig. 2. System implementation of a Hilbert transformer exploiting a microring-based comb source. VNA is a Vector Network Analyzer, WS= wavelength selective switch, VOL=variable optical attenuator.

device – first, a reference for time 'zero' is required, as illustrated in Fig. 1(a). Secondly, negative tap coefficients are required.

Practical implementations of photonic transversal filters are often realized by assigning filter taps to different optical wavelengths. The time delay between filter taps (wavelengths) can be achieved by using a dispersive medium such as a long optical fiber [12] or a chirped grating [15]. In previous demonstrations of a Hilbert transformer using photonic transversal filtering, discrete filter taps were realized by employing an array of discrete continuous-wave laser sources. However, this approach limits the number of filter taps to only four, resulting in less than a 3-octave bandwidth. In this work, we use a wide spectrum integrated comb source based on a high-Q microring resonator, thus allowing a much higher number of filter taps, in turn significantly broadening the RF bandwidth.

Figure 2 illustrates the experiment setup, which consists of three main sections: frequency comb generation, comb shaping, and photonic RF filtering. A continuous-wave (cw) tunable laser (Agilent 8160A), amplified by a high power Erbium-doped fiber amplifier (EDFA1), was used as the pump source for the high-Q microring resonator. The equally spaced comb lines produced by the ring resonator were then shaped according to the required tap coefficients using a reconfigurable filter (a Finisar S400 waveshaper), having a much smaller resolution than the comb line spacing. The waveshaper also split the comb into two paths, which were connected to the photonic RF filter section. Comb lines corresponding to positive tap coefficients were routed to one of the output ports of the waveshaper, while comb lines corresponding to negative tap coefficients were sent to the second port.

The two outputs of the waveshaper were connected to two inputs of a 2x2 balanced Mach-Zehnder modulator (MZM), biased at quadrature, where the comb lines were modulated by the RF signal. One group of comb lines was modulated on the positive slope of the MZM transfer function while the second group was modulated on the negative slope. This allowed both negative and positive tap coefficients to be realized with a single MZM. The output of the MZM was then passed through 2.122 km of single-mode fiber (SMF) that acted as a dispersive element to delay the different filter taps. The dispersed signal was then amplified by a second fiber amplifier (EDFA2) to compensate for loss, and filtered in order to separate the comb from the signal at the pump wavelength in order to produce the system $0^{o}$ phase reference. The second path was used as the $90^{o}$ phase signal. The signal path was time-shifted with a variable optical length (VOL) so that the reference could be positioned, in time, exactly at the middle of the filter taps, as illustrated in Fig. 1(a). The optical signals were finally detected by photodiodes to regenerate output RF signals. The system RF frequency response was then measured with an RF vector network analyzer (VNA).

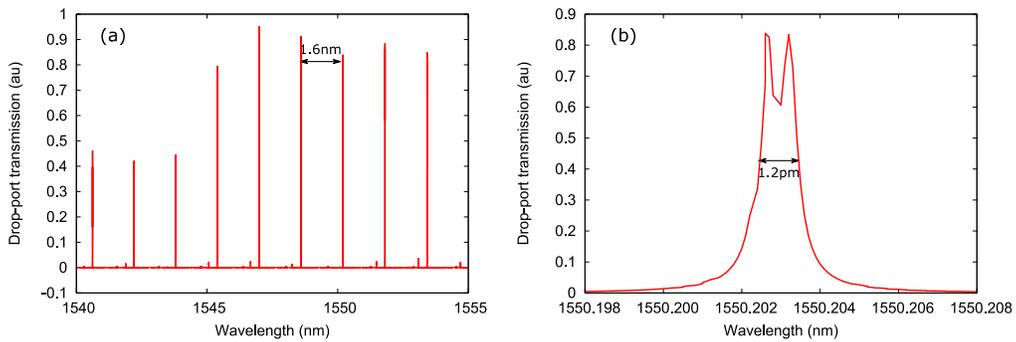

Fig. 3. (a) The wavelength response of the drop-port of the microring resonator; (b) zoom-in of a resonance close to 1550nm showing the resonance width (FWHM) of 1.2 pm (150 MHz)

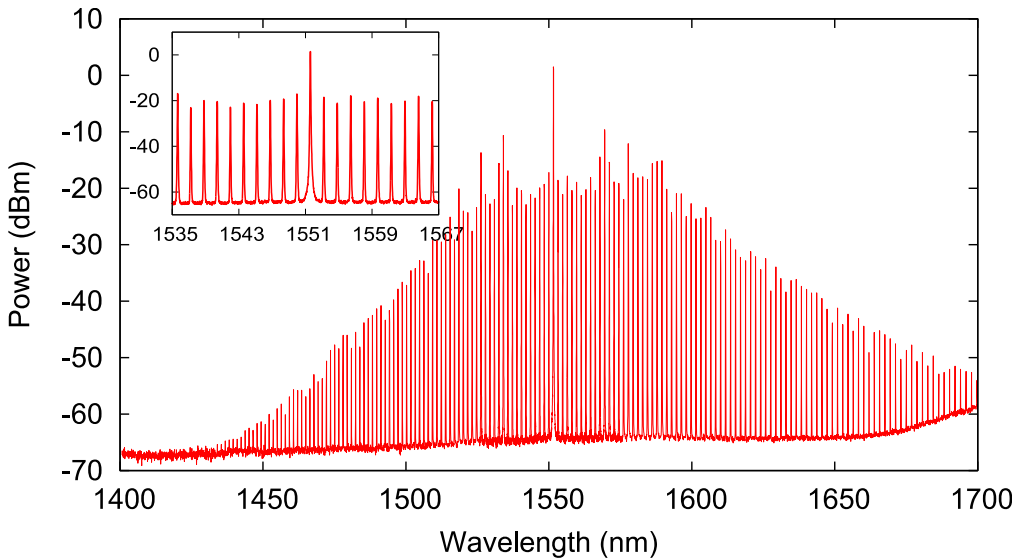

Fig. 4. Spectrum of the comb generated from the microring, measured by an OSA with a resolution of 0.5nm. Inset: Zoom-in of the spectrum around the pump wavelength.

## 3. Frequency comb source

The frequency comb source was achieved using a high-Q microring resonator [25]. The microring was fabricated in a high-index contrast doped-silica glass platform using CMOS-compatible processes [24, 27-30]. The waveguide width and height were both 1.5 μm while the radius of the ring was 135 μm, yielding a free spectral range (FSR) of 1.6 nm, or 200 GHz, at 1550 nm. The wavelength transmission response at low power, measured at the ring drop-port is shown in Fig. 3(a). Fig. 3(b) shows a high-resolution plot of the drop port transmission response at one of the resonances. A split resonance is evident due to the contra-directional coupling caused by surface roughness [31, 32]. From the resonance width, the quality factor of the microring was estimated to be $Q = 1.3 \times 10^6$. The maximum power at the drop-port output at resonance was measured to be approximately 14 dB below the input power.

To generate the comb, the pump laser wavelength was tuned to one of the resonances near 1550 nm. When the pump power was 0.5 W, the spectrum measured at the drop port exhibited a broad frequency comb. Fig. 4 shows the spectrum of the comb, spanning more than 250nm, as measured by an optical spectrum analyzer (OSA, with a resolution of 0.5 nm). The inset of Fig. 4 shows a high-resolution scan of the measured comb spectrum around the pump wavelength. The spacing of the comb lines corresponded to the FSR of the microring.

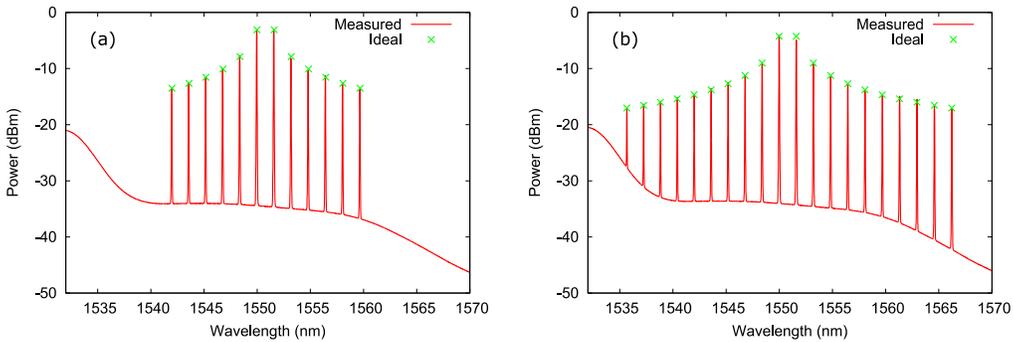

Fig. 5. EDFA2 output showing the weight of each tap for: (a) 12 tap filter, and (b) 20 tap filter

The power of the comb line at the pump wavelength was about 20 dB higher than the neighboring lines.

The generated comb was a "Type II" comb [33] – where the coherence between the comb lines was not particularly high [34]. However, for the photonic Hilbert transformer demonstrated here, the optical signals from different taps were detected incoherently by the photodiode and so a high degree of coherence between comb lines is not required in order to achieve high RF performance.

## 4. Experiment

### 4.1. Comb shaping

Since the spectrum of the comb generated by the microring did not follow the hyperbolic function required for the impulse response of a Hilbert transformer [Fig. 1(a)], it was necessary to shape the comb in order to achieve the required tap coefficients. Each individual comb line was selected and shaped to the desired power level by the waveshaper. This could be easily achieved since its resolution (10 GHz) was much smaller than the comb spacing (200 GHz). The normalized power of each comb line, needed to achieve a Hilbert Transform, is given by:

$$p_n = \frac{1}{\pi |n - N/2 + 0.5|} \qquad (1)$$

where *N* is the number of comb lines, or filter taps, used in the filters, and *n = 0, 1, 2, ..., N-1* is the comb index.

Figures 5(a) and 5(b) show the powers of all the comb lines measured at the output of the EDFA2 using an OSA for the 12 tap and 20 tap filter cases, respectively. The target powers at all wavelengths are also shown in Figs. 5(a) and 5(b) as green crosses. The waveshaper was successfully used to shape the powers of all comb lines to within +/-0.5 dB of the target powers. Unused comb lines were attenuated below the noise floor.

*4.2. System RF response*

Once the comb lines were attenuated in order to provide the correct tap coefficients of the impulse response associated with a Hilbert transform, the system RF frequency response was then characterized. A vector network analyzer (VNA) was used to measure the system RF amplitude and phase frequency response. First, the VNA was calibrated with respect to the reference output $RF_{REF}$ and then the signal output $RF_{SIG}$ was measured with the calibrated VNA.

Figure 6 (a) shows the measured RF amplitude frequency response of the photonic Hilbert transform filters for 12, 16 and 20 taps, respectively, which all exhibit expected behavior.

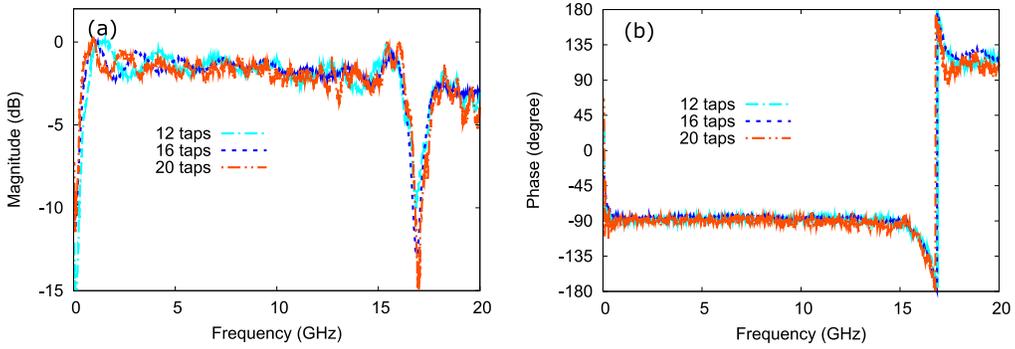

Fig. 6. Measured system RF frequency response for different number of filter taps: (a) amplitude; and (b) phase response

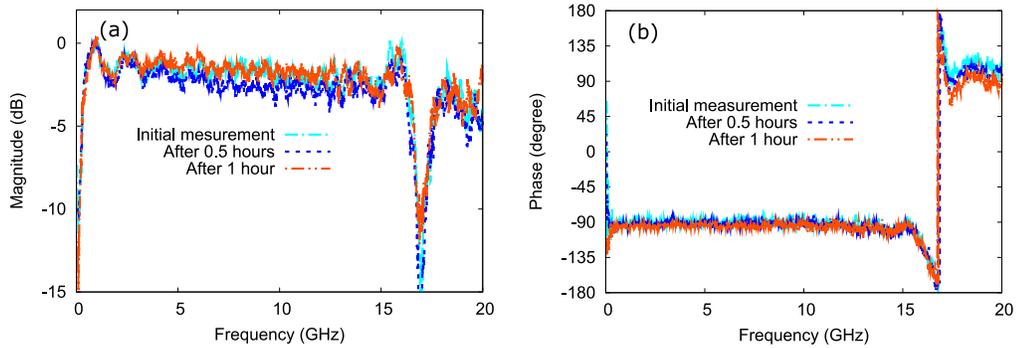

Fig. 7. Measured system RF frequency response of the 20 tap filter measured at different times: (a) amplitude; and (b) phase response

All three filters have the same null frequency at 16.9 GHz, corresponding to the tap spacing of $\Delta t = 1/f_c = 59$ ps. This spacing matches the difference in delay between the comb lines, equal to the ring $FSR = 1.6$ nm, produced by propagation through a 2.122 long SMF fiber with a dispersion parameter $D = 17.4$ ps/(nm km). The null frequency could be controlled by using a different fiber length to adjust the tap spacing.

All filters show < 3 dB amplitude ripple. As predicted in Fig. 1(c), increasing the number of filter taps increases the filter bandwidth. With a 20 tap filter, the Hilbert transformer exhibited a 3 dB bandwidth extending from 16.4 GHz down to 0.3 GHz, corresponding to more than 5 octaves. It is possible to increase this bandwidth further by using more comb lines in the filter. In our experiment, only a small portion of the generated comb spectrum was actually used to realize the filter taps. The number of filter taps that could be achieved was actually limited by the bandwidth of the waveshaper and the gain bandwidth of the optical amplifier (EDFA2). If desired, it would also be possible to reduce the amplitude ripple within the pass-band by apodizing the tap coefficients from the ideal hyperbolic function [12].

Figure 6(b) shows the measured phase response of filters with different numbers of taps, showing very similar responses. Each shows a relatively constant phase of near -90° within the pass-band. There are some deviations from the ideal -90° phase at frequencies close to zero and particularly for the null frequency $f_c = 16.9$ GHz. The reasons for the phase errors at the band edges are discussed in the following section.

To assess the stability of the system, the RF response was measured at different times. Figs. 7(a) and 7(b) show the RF amplitude and phase frequency responses of the 20 tap filter measured immediately after the system was set up, after 30 minutes and then after 1 hour. It can be seen that there is a small variation of up to 1 dB in the RF amplitude frequency response when the system was characterized at these different times, whereas the system shows a similar phase response over the pass band except a small phase variation at the band edges. The small fluctuation in the system response can be attributed mainly to the drift in the bias of the Mach- Zehnder modulator. When the modulator bias drifts the filter tap coefficients will depart from the ideal values resulting in a change in the system RF response. An active bias-controller can in principle be used to minimize the effects of bias drift.

## 5. Effects of imperfections on RF phase response

The theoretical results of Fig. 1(d) suggest that a Hilbert transformer based on transversal filtering has a frequency independent phase response from the zero frequency up to the null frequency regardless of the number of filter taps. Indeed, we obtain [Fig. 6(b)] a frequency independent -90° phase response within the pass-band of the photonic Hilbert transformer. However, we do observe phase errors at the band-edges, especially at the null frequency, which can be caused by imperfections in the device configuration. In order to achieve a frequency independent phase response within the pass-band, the impulse response of a Hilbert transformer must be a perfectly anti-symmetric function with equal tap spacing, as illustrated in Fig. 1(a). Any error in the power of the comb lines, the presence of higher order dispersion terms in the dispersive fiber, or the modulator chirp can all contribute to deviations of the filter impulse response from the ideal case.

To investigate the effects of imperfections, the system RF response was simulated using VPI simulator software [35] for different tap coefficients, for variations in the third order dispersion (TOD) of the fiber, and for modulator chirp.

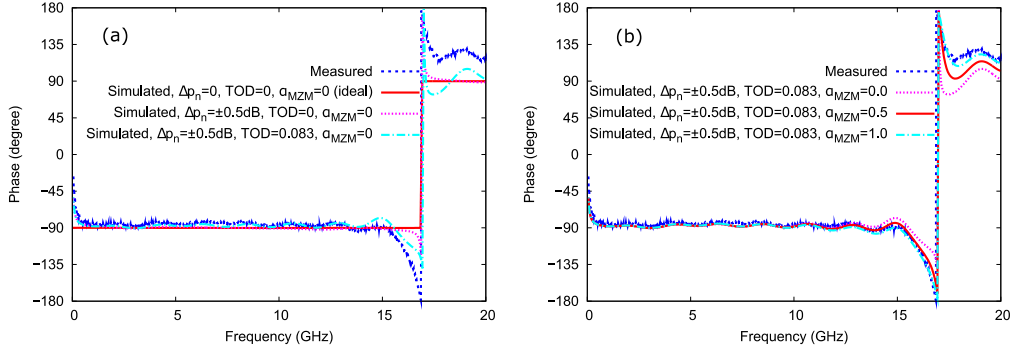

Fig. 8. The measured and simulated phase responses of the 16 tap filters. Simulated results showing the effects of: (a) tap coefficient error $\Delta p_n$, third order dispersion in the fibre delay line, and (b) the modulator chirp $\alpha_{MZM}$.

Figure 8(a) shows the simulated phase response of the 16 tap fiber for different values of the tap coefficients ($\Delta p_n$) and TOD, while keeping the modulator chirp zero ($\alpha_{MZM} = 0$). The measured phase response is also plotted in Fig. 8(a) for easy comparison. It can be seen from Fig. 8(a) that, apart from a very small phase ripple within the pass-band, the deviation of the tap coefficients from the ideal values mainly causes phase degradation at the band edges. However, unlike our experimental observations (Fig 6(b)), phase errors caused by tap coefficient variations are the same at both band edges. The deviation of the tap coefficients from the ideal values causes an imbalance between the positive and negative tap coefficients. This degrades the perfect anti-symmetry of the impulse response, which in turn results in phase degradation at the band edges.

The simulation results in Fig. 8(a) show that when fiber TOD corresponding to a dispersion slope of 0.083 ps/(nm$^2$ km) near 1550 nm is taken into account, the phase ripple increases as the RF frequency increases. Non-zero TOD also causes a large phase error at the null frequency $f_c$. It has been known that fiber TOD introduces a second-order phase in the filter taps [36, 37], thus resulting in non-uniform tap delays between adjacent filter taps [26, 36].

In addition to the desired intensity modulation, most practical MZMs also induce phase modulation – a phenomenon call modulator chirp [38]. The effect of modulator chirp on the filter phase response is shown in Fig. 8(b), where it is seen that it also contributes to phase degradation at the null frequency $f_c$. This modulator chirp introduces additional non-uniformity in the spacing between filter taps.

It is clear from the simulation results of Fig. 8 that if the tap coefficient ripple, the fiber TOD and modulator chirp are all taken into account, we achieve a simulated phase response that matches the measured data very well. In order to minimize the phase error, the deviation of the tap coefficients from the ideal values needs to be reduced, and the tap spacing non-uniformity caused by the fiber TOD and the modulator chirp should be minimized. It should be possible to reduce the tap coefficient deviation by minimizing the amplitude shaping error associated with the waveshaper through feedback control. In [12], near perfect tap spacing uniformity was achieved by carefully adjusting the wavelength separation of all lasers. The spacing between the comb lines used in this work was fixed by the ring FSR. Therefore, to reduce the tap spacing non-uniformity, the fiber TOD should be compensated, a low chirp modulator used and the phase modulation due to modulator chirp corrected. Each of these conditions could be satisfied by using a second reconfigurable filter in the setup with a tailored phase profile to cancel out the second-order phase in the filter taps [26, 36] as well as the phase modulation due to modulator chirp. A discrete Bragg grating could also be used instead of a long fiber to minimize TOD effect [15].

## 6. Discussion

We have shown that an integrated optical comb source can be effectively used to provide numerous, high quality optical taps for a microwave photonic transversal filter, thus allowing us to demonstrate a very wide bandwidth RF Hilbert transformer with a 3 dB bandwidth of over 5 octaves from as low as 0.3 GHz to 16.9 GHz. It is extremely difficult to match this performance using electronic or other photonic techniques. In this work, only a small number of the available comb lines from the integrated comb source were utilized to realize the filter taps. This was limited by the finite bandwidth of the configurable filter used to shape the comb spectrum as well as the optical fiber amplifiers. Reducing loss in order to potentially eliminate the amplifier, as well as using a wider bandwidth configurable filter, will both allow more comb lines to be used, resulting in an even broader RF bandwidth. Further improvements to the filter ripple and response near the band edges can be achieved through apodization and compensation of imperfections in the modulation and transmission system. Since the integrated comb source can generate many more comb lines than the number of filter taps, it is possible to realize multiple parallel filters using only a single comb source, further reducing the device complexity.

Although our device is still relatively bulky due to the discrete components that were employed, such as the waveshaper, the integrated nature of the comb source has significant potential to reduce the system complexity by combining many different functions on an integrated chip. In addition to enabling high quality comb sources, the high nonlinearity of the Hydex platform as well as that of other integrated comb source platforms [24] is also attractive for additional on-chip signal processing functions. For example, four wave mixing in highly nonlinear waveguides and ring resonators [24] can be combined with a Hilbert transformer to realize devices capable of instantaneous frequency measurements [14].

## 7. Conclusions

We demonstrate a wideband Hilbert transformer using a transversal filtering scheme. An integrated parametric frequency comb source generated by a CMOS-compatible nonlinear, high-Q microring resonator is exploited in order to achieve multi-tap transversal filters. The wide spectral width and large frequency spacing of the integrated comb source allows a large number of high quality filter taps without any increase in the system complexity, thus achieving a much wider RF bandwidth than what can be typically obtained with standard microwave circuits. We achieve more than a 5-octave bandwidth while maintaining a nearly frequency independent quadrature phase over the pass-band. Our approach also has the potential to achieve full integration on a chip, which could have a significant impact on signal processing systems for many applications including radar detection, imaging and communications.